\documentclass[sn-mathphys]{sn-jnl}

\jyear{2021}%

\theoremstyle{thmstyleone}%
\theoremstyle{thmstyletwo}%

\theoremstyle{thmstylethree}%

\begin{document}

\title[Evaluation of the eddy diffusivity in a pollutant dispersion model in the planetary boundary layer]{Evaluation of the eddy diffusivity in a pollutant dispersion model in the planetary boundary layer}


\author*[1]{\fnm{} \sur{A. Goulart}}\email{antonio.goulart@furg.br}

\author[1]{\fnm{} \sur{J. M. S. Suarez}}\email{julian.sejje@furg.br}
\equalcont{These authors contributed equally to this work.}

\author[1]{\fnm{} \sur{M.J. Lazo}}\email{matheuslazo@furg.br}
\equalcont{These authors contributed equally to this work.}

\author[1]{\fnm{} \sur{J.C. Marques}}\email{joicemtm@furg.br}
\equalcont{These authors contributed equally to this work.}

\affil[1]{\orgdiv{Universidade Federal do Rio Grande}, \orgname{IMEF}, \orgaddress{ \city{Rio Grande}, \postcode{96.201-900}, \state{RS}, \country{Brazil}}}


\abstract{In this work, eddy diffusivity is derived from the energy spectra for the stable and convective regimes in the planetary boundary layer. The energy spectra are obtained from a spectral model for the inertial subrange that considers the anomalous behavior of turbulence. This spectrum is expressed as a function of the Hausdorff fractal dimension. The diffusivity eddies are employed in a classical Eulerian dispersion model, where the derivatives are of integer order and in fractional dispersion model, where the derivatives are of fractional order. The eddy diffusivity proposed considers the anomalous nature of geophysical turbulent flow. The results obtained with the fractional  and classic dispersion models using the eddy diffusivity proposed is satisfactory when compared with the experimental data of the Prairie Grass and Hanford experiments in a stable regime, and the Copenhagen experiment in a convective regime.}




\maketitle

\section{Introduction}\label{sec1}

The dispersion of pollutants in the earth's atmosphere is a permanent source of defiant problems due to its physical complexity. An example of a non\-trivial problem is the description of the dispersion of the pollutant in the stable and convective boundary layer.
 Several works describe that the turbulent activity in fully developed turbulence seems to be concentrated in a fractal \cite{Mendelbrot1974, Sreenivasan1986, Procaccia1992, Mandelbrot1995, Mazzi2004, Thormann2014, Bai2015}  and that anomalous transport processes are sensitive to the fractal nature of turbulence \citep{Hentschel1983, Metzler2000, Metzler2004}.  In common diffusion, the mean-square displacement increases linearly with time, unlike in anomalous diffusion, where the mean-square displacement is not linear. The anomalous diffusion is closely connected with the failure of the central limit theorem due to a sparse distribution or long-range correlations. Actually, the anomalous diffusion is related to the more general L\'{e}vy--Gnedenko theorem, which generalizes the central limit theorem for si\-tuations where not all moments exist \cite{Metzler2000}. Historically, anomalous diffusion was first observed in nature in the dispersion of contaminants phenomenon. In 1926, Richardson measured the increase in the width of plumes of smoke generated from point sources located in a  turbulent velocity field \cite{Richardson1926}). Based on his observations, Richardson speculated that the velocity of turbulent air, which has a non-differentiable structure, can be approximately described by the Weierstrass function. This was motivated partly by the observation that the width of smoke plumes grows with $t^{\alpha}$ ($\alpha\geq 3$), unlike common diffusion where $\alpha=1$. Furthermore, the non-differentiable behavior of the width growth of plumes generated from a point source is directly related to the fractal structure of the turbulent velocity field, where the fluctuation's size scales are, in many cases, very large compared to the average scale. In this context, traditional differential equations do not adequately describe the problem of turbulent diffusion because usual derivatives are not well defined in the non-differentiable behavior introduced by turbulence.
 Goulart et al. \cite{Goulart2017} and Goulart et al. \cite{Goulart2020} proposed a contaminant dispersion model in the planetary boundary layer that does not solve the advection-diffusion equation as expressed traditionally but modifies the mathematical structure of this equation to represent more realistically the evolution in space of the concentration of contaminants dispersed in a turbulent flow. In this sense, fractional operators were used in the equations that govern the distribution of contaminants in the planetary boundary layer. The use of fractional calculus in the modeling of turbulent diffusion is justified by the non-differentiable behavior of the problem and by the presence of anomalous diffusion. In the works of Goulart et al. \cite{Goulart2017} and Goulart et al. \cite{Goulart2020}, the eddy diffusivity used at the dispersion model is constant.
 In a recent work by Goulart et al. \cite{Goulart2022},  we developed a model for the velocity spectra that considers the anomalous behavior of the turbulent flow, observed in some particular cases in Earth's atmosphere, as is the case with the stable boun\-dary layer. In the work of Goulart et al. \cite{Goulart2022}, to deduce the velocity spectra that considers the anomalous behavior of the turbulent flow, the $\beta$-model was used. The $\beta$-model assumes that the standard Kolmogorov phenomeno\-logy is valid only in active turbulence regions, and it proposes an expression for the turbulent velocity spectra in the inertial subrange that is a function of the Hausdorff fractal dimension \cite{Frisch1978,Novikov1964}. From this idea, expressions are obtained for the components of the turbulent velocity spectra that describe the turbulence that exists in geophysical turbulence above the ocean and in the stable boundary layer where intermittent turbulence occurs. With these spectra, eddy diffusi\-vity is obtained. In the present work, we extend the work of Goulart et al. \cite{Goulart2022}, developing a fractal velocity spectrum for the convective stability regime in the planetary boundary layer. From this spectrum, eddy diffusivity expressed as a function of the Hausdorff fractal dimension is obtained. Furthermore, a new fractal eddy diffusivity for the stable regime is obtained from the velocity spectrum suggested by Goulart et al. \cite{Goulart2022}. These eddy diffusivities are used in two  Eulerian dispersion models to estimate the concentration of contaminants in the stable and convective boundary layer. First, in the classical dispersion model, it is solved for the advection-diffusion equation. After, in the model proposed by Goulart et al. \cite{Goulart2017}, where fractional operators are introduced in the equation that governs the distribution of contaminants in the atmosphere. The eddy diffusivity obtained as a function of the fractal dimension more adequately describes a geophysical flow, where anomalous diffusion occurs.
 In this work, the spatial distribution of the concentration of a non-reactive pollutant in a stable boundary layer obtained with these models are compared with experimental data from the Prairie Grass and Hanford experiments, and for the convective case, the model is compared with experimental data from the Copenhagen experiment.    \\
The paper is organized in the following way: In Section II, the parameterization of turbulence in the planetary boundary layer is presented. In Section III, the dispersion models are presented.  A numerical comparison with the models and experimental data is done in Section IV. Finally, Section V presents the conclusions.

\section{Turbulence parameterization in planetary boundary layer}\label{sec2}
\subsection{{ Derivation of a vertical asymptotic fractal eddy diffusivity}}

 To obtain an asymptotic (time-independent) vertical eddy diffusivity, we will consider the expression for the eddy diffusivity obtained by Batchelor \cite{Batchelor1949} and used in several works found in the literature \cite{Batchelor1949, Degrazia2000, Goulart2004}:

\begin{equation}
K_{\gamma}(x,z)=\frac{\sigma_{j}^{2}\beta_{j}}{2\pi} \int_{0}^{\infty} E_{j}(n)\sin{\big( \frac{2\pi x n}{u \beta_{j}}\big) }\frac{dn}{n}, \;\;\; \gamma = x,y,z, \;\;\;j=u,v,w.
\label{eq 1}\end{equation}
where $x,y,z$ are the components of the Cartesian coordinate system in which the x-axis coincides with the direction of the average wind and $z$ is the vertical axis, $u = u(z)$ is the mean wind velocity component in the longitudinal direction and $E_{j}(n)$ is the value of the Eulerian spectrum of energy normalized by the Eulerian velocity variance,

\begin{equation}
E_{j}(n)=\frac{S_{j}(n)}{\sigma_{j}^{2}}.
\label{eq 2}\end{equation}

 The Eq. \eqref{eq 1} describes the eddy diffusivity ($K_{\gamma}(x,z)$) from a Lagrangian perspective, but the turbulent velocity spectra $S_{j}$ is an Eulerian physical quantity. The $\beta_{j}$ parameter relates these two perspectives \cite{Hanna1981}.

For large diffusion travel times, the asymptotic behavior of the integral in Eq. \eqref{eq 1} can be considered

\begin{equation}
K_{\gamma}(x,z)=\frac{\sigma_{j}^{2}\beta_{j}}{2\pi} E_{j}(0)\int_{0}^{\infty} \sin{\big( \frac{2\pi x n}{u \beta_{j}}\big) }\frac{dn}{n}.
\label{eq 3}\end{equation}
 The integral in Eq. \eqref{eq 3} can be calculated analytically. Its value is $\pi/2$. In this case, the eddy diffusivity for large travel time assumes the form,

\begin{equation}
K_{\gamma}(z) = \frac{1}{4}\sigma_{j}^{2}\beta_{j} E_{j}(0).
\label{eq 4}\end{equation}
 The parameter $\beta_{j}$ can be found in the literature \cite{Wandel1962, Angell1974, Hanna1981, Degrazia1998} in the following form

\begin{equation}
\beta_{j}= a \frac{u}{\sigma_{j}},
\label{eq 5}\end{equation}
where $a$ is a constant, $\sigma_{j}$ is the Eulerian standard deviation of the turbulent wind velocity and $u$ is the average wind velocity. Substituting Eq. \eqref{eq 5} in Eq. \eqref{eq 4} we can write the following expression for the asymptotic eddy diffusivity:

\begin{equation}
K_{\gamma} (z) = \frac{a}{4}u E_{j}(0)\sigma_{j}.
\label{eq 6}\end{equation}

 In the next section, we will obtain the vertical asymptotic eddy diffusivity for the stable and convective stability regimes existing in the planetary boundary layer.

\subsubsection{Asymptotic vertical eddy diffusivity for the stable boundary layer}
To obtain an expression for the asymptotic vertical eddy diffusivity in the stable boundary layer from Eq. \eqref{eq 6} ($\gamma = z$ and $ j=w$), we will calculate $E_{w}(0)$ and $\sigma_{w}$ from the vertical velocity spectrum obtained by Goulart et al. \cite{Goulart2022},
\begin{equation}
S_{w}(n)  = \frac{\frac{3}{5-3D} (2 \pi)^{D-\frac{5}{3}}C_{w} \kappa^{-\frac{2}{3}}\ell_{w}^{D-1} z^{1-D} \Phi_{\epsilon}^{\frac{2}{3}}\frac{z}{u}u_{*}^{2}}{[(f_{m})_{w}]^{\frac{8}{3}-D} \big[1+\frac{3}{5-3D}(\frac{f}{(f_{m})_{w}})^{\frac{8}{3}-D}\big]}
\label{eq 7}\end{equation}
where $\kappa=0.4$ is the von Karman constant, $C_{w}=2/3$, $u_{*}$ is the friction velocity, $\ell_{w}$ is the vertical length scale, $\Phi_{\epsilon}$ is the dimensionless dissipation rate, $(f_{m})_{w}$ is the frequency of the spectral peak, and $D$ is the Hausdorff dimension.
Replacing $n=0$ in Eq. \eqref{eq 7} we get,
\begin{equation}
S_{w}(0)  = \frac{3}{5-3D} (2 \pi)^{D-\frac{5}{3}}C_{w} \kappa^{-\frac{2}{3}}\ell_{w}^{D-1} z^{1-D} \Phi_{\epsilon}^{\frac{2}{3}}\frac{z}{u}[(f_{m})_{w}]^{D-\frac{8}{3}} u_{*}^{2}.
\label{eq 8}\end{equation}

The variance of the vertical turbulent wind velocity ($\sigma_{w}^{2}$) is obtained by integrating the velocity spectrum given by Eq. \eqref{eq 7} of the $n=0$ to $n=\infty$,

\begin{equation}
\begin{array}{lll}
 \sigma_{w}^{2} =& (2 \pi)^{D-\frac{5}{3}}\Gamma[1+\frac{3}{8-3D}]\Gamma[1-\frac{3}{8-3D}]\big( \frac{3}{5-3D}\big)^{\frac{3D-5}{3D-8}}C_{w} \kappa^{-2/3}\ell_{w}^{D-1} \\
  & \times \; z^{1-D} \Phi_{\epsilon}^{\frac{2}{3}}[(f_{m})_{w}]^{D-\frac{5}{3}}u_{*}^{2}
\end{array}
\label{eq 9}\end{equation}
where $\Gamma[.]$ is the Gamma Function.\\
Replacing Eqs. \eqref{eq 8} and \eqref{eq 9} in Eq. \eqref{eq 2} (considering $n=0$), we get,

\begin{equation}
E_{w}(0) = \big( \frac{3}{5-3D}\big)^{-\frac{3}{3D-8}}\frac{z}{u\;(f_{m})_{w}}\Gamma[1+\frac{3}{8-3D}]\Gamma[1-\frac{3}{8-3D}].
\label{eq 10}\end{equation}

Finally, replacing $E_{w}(0)$ (given by Eq. 10) and the Eulerian standard deviation of the turbulent wind velocity $\sigma_{w}$ obtained from Eq. \eqref{eq 9}, we obtain the following expression for the vertical asymptotic eddy diffusivity in the stable boundary layer:

\begin{equation}
K(z) = \frac{a(2 \pi)^{\frac{3D-5}{6}}\big( \frac{3}{5-3D}\big)^{\frac{3D-11}{2(3D-8)}}C_{w}^{\frac{1}{2}}\kappa^{-\frac{1}{3}}\ell_{w}^{\frac{D-1}{2}} z^{\frac{3-D}{2}} \Phi_{\epsilon}^{\frac{1}{3}}[(f_{m})_{w}]^{\frac{3D-11}{6}}u_{*}}{4\big(\Gamma[1+\frac{3}{8-3D}]\Gamma[1-\frac{3}{8-3D}]\big)^{\frac{1}{2}}}.
\label{eq 11}\end{equation}

In Eq. \eqref{eq 11} the following expression is used for the dimensionless dissipation rate \cite{Degrazia1998},

\begin{equation}\label{eq 12 }
\Phi_{\epsilon}=1.25(3.7 {z}/{\Lambda})
\end{equation}
where,
\begin{equation}\label{eq 13 }
\Lambda = L (1-{z}/{h})^{\frac{5}{4}},
\end{equation}
is the local Obukhov length, $L$ is the Obukhov length, and $h$ is the stable boundary-layer height. For vertical length scale $\ell_{w}$ and vertical reduced frequency of the spectral peak $(f_{m})_{w}$ the following expressions \cite{Degrazia2000}:
\begin{equation}\label{eq 14 }
\ell_{w}=\frac{0.27z }{1 + 3.7 \frac{z}{\Lambda}}
\end{equation}
and
\begin{equation} \label{eq 15 }
(f_{m})_{w}=0.33(1+3.7\frac{z}{\Lambda}) .
\end{equation}
The equation used to calculate the mean wind velocity profile is described by a power law \cite{Panofsky1984}
\begin{equation} \label{eq 16}
u= u_{0}\big(\frac{z}{z_{0}}\big)^{p}.
\end{equation}
In this work, the value $p=0.09$ was used for the Copenhagen experiment, $p=0.60$ for the Hanford experiment and $p=0.10$ for the Prairie Grass experiment (stable case).
\subsubsection{Asymptotic eddy diffusivity for the convective boundary layer}

To obtain an expression for the asymptotic convective eddy diffusivity from Eq. \eqref{eq 6}, we need to obtain an expression for the velocity spectrum in the convective boundary layer. To do this, we will consider the $\beta$-model, similar to what was done by Goulart et al. \cite{Goulart2022} for the stable spectrum. According to the $\beta$ model, the fractal kinetic energy density in the inertial subrange is given by \cite{Novikov1964, Frisch1978, Goulart2022},

\begin{equation}
 nS_{j}(n)=C_{j} \ell_{j}^{-(1-D)} \langle \epsilon \rangle ^{\frac{2}{3}} (\frac{z}{2 \pi})^{\frac{5-3D}{3}} f^{\frac{3D-5}{3}}, \;\;\;\;j=u,v,w
\label{eq 17} \end{equation}
where, $C_{j}$ is the Kolmogorov constant and $\langle \epsilon \rangle$ dimensionless dissipation rate.

The spectrum given by Eq. \eqref{eq 17} is valid in the inertial subrange. To obtain a valid spectrum at all frequencies, we will consider the fitting curve for the convective spectrum below \cite{Sorbjan1989}

\begin{equation}
\frac{nS_{j}(n)}{w_{*}^{2}}= \frac{A_{j} f^c}{(1+B_{j} f^{a})^{b}},
\label{eq 18} \end{equation}
where $A_{j}$, $B_{j}$, $a$, $b$ and $c$ are constants and $w_{*}$ is the convective velocity scale. From some criteria,  it is possible to obtain the value of the constants $A_{j}$, $B_{j}$, $a$, $b$ and $c$. The first criterion is that the function in Eq. \eqref{eq 18} must be consistent with the law for the inertial subrange in Eq. \eqref{eq 17}. For large frequencies, the term $B_{j} f^{a} \gg 1$  in the denominator of Eq. \eqref{eq 18}.  In this case,  the Eq. \eqref{eq 18} tends to present as
\begin{equation}
\frac{nS_{j}(n)}{w_{*}^{2}} \rightarrow  f^{c-ab}.
\label{eq 19} \end{equation}
Considering the large frequencies in Eq. \eqref{eq 17} and comparing the potencies of $f$ in Eqs. \eqref{eq 17} and \eqref{eq 19}, we obtain
\begin{equation}
c-ab = \frac{3D-5}{3}.
\label{eq 20}
\end{equation}
Another criterion arises from the need for the position of the maximum spectrum $(f_{m})_{j}$ to coincide with the maximum of the function in Eq. \eqref{eq 18}. Deriving the right side of the Eq. \eqref{eq 18} with respect to $f$ and equalizing to zero we get:
\begin{equation}
(f_{m})_{j}= \big[\frac{c}{B_{j}(ab-c)}\big]^{\frac{1}{a}}.
\label{eq 21} \end{equation}
For  large frequencies in Eq. \eqref{eq 18} ($B_{j} f^{a} \gg 1$), we get $nS_{j}(n) = ({A_{j}}/{(B_{j})^b})f^{c-ab}w_{*}^{2}$. Equating this expression to Eq. \eqref{eq 17} and considering Eq. \eqref{eq 20}, we obtain
\begin{equation}
A_{j} = (\frac{3}{5-3D})^{\frac{8}{3}-D}\frac{C_{j}\ell_{j}^{D-1} }{w_{*}^{2} } (\frac{z}{2 \pi})^{\frac{5-3D}{3}} \langle \epsilon \rangle ^{\frac{2}{3}} B_{j}^{b} \big..
\label{eq 22}
\end{equation}
From the  Eqs. \eqref{eq 20}, \eqref{eq 21} and \eqref{eq 22}, consider $a = c = 1$ (Sorbjan 1989) and $\psi_{\epsilon} = {\langle \epsilon \rangle  z_{i}}/{w_{*}^{3}}$, where $\psi_{\epsilon}$ is the nondimensional molecular dissipation rate function in the convective conditions, we get
\begin{equation}
A_{j}=(\frac{3}{5-3D})^{\frac{8}{3}-D} (2 \pi)^{D-\frac{5}{3}}C_{w} \ell_{w}^{D-1}  (\frac{z}{z_{i}})^{\frac{2}{3}}z^{1-D}\psi_{\epsilon}^{\frac{2}{3}}[(f_{m})_{j}]^{D-\frac{8}{3}}
\label{eq 23}\end{equation}
and
\begin{equation}
B_{j} = \frac{3}{(5-3D)[(f_{m})_{j}]}.
\label{eq 24}\end{equation}
By replacing Eqs. \eqref{eq 23} and \eqref{eq 24} in Eq. \eqref{eq 18} ($a = c = 1$ and $j=w$), we obtain the expression for the vertical fractal turbulent velocity spectra in the convective boundary layer
\begin{equation}
\frac{nS_{w}(n)}{w_{*}^{2}}  = \frac{(\frac{3}{5-3D})^{\frac{8}{3}-D} (2 \pi)^{D-\frac{5}{3}}C_{w} \ell_{w}^{D-1} (\frac{z}{z_{i}})^{\frac{2}{3}}z^{1-D} \psi_{\epsilon}^{\frac{2}{3}}f}{[(f_{m})_{w}]^{\frac{8}{3}-D} \big[(1+\frac{3}{5-3D}\frac{f}{(f_{m})_{w}})^{\frac{8}{3}-D}\big]}.
\label{eq 25}\end{equation}

Considering the relation $f=\frac{nz}{u}$ in Eq. \eqref{eq 25} and calculating the spectrum at $n=0$, we obtain,

\begin{equation}
S_{w}(0)=(\frac{3}{5-3D})^{\frac{8}{3}-D} \frac{C_{w}}{u}(2 \pi)^{D-\frac{5}{3}} \ell_{w}^{D-1} (\frac{z}{z_{i}})^{\frac{2}{3}}z^{1-D} \psi_{\epsilon}^{\frac{2}{3}}[(f_{m})_{w}]^{\frac{8}{3}-D}
\label{eq 26}\end{equation}
where $C_{w}=2/3$, $w_{*}$ is the convective velocity scale, $z_{i}$ is the top of the unstable boundary layer height, $(f_{m})_{w}$ is the normalized frequency of the spectral peak regardless of stratification and $\psi_{\epsilon}$ is the nondimensional molecular dissipation rate function, defined by $\psi_{\epsilon} = \epsilon z_{i}/w_{*}^{3}$.

The variance of the vertical turbulent wind velocity ($\sigma_{w}^{2}$) is obtained by integrating the velocity spectrum given by Eq. \eqref{eq 25} of the $n=0$ to $n=\infty$,
\begin{equation}
\sigma_{w}^{2} =\frac{2}{3}(\frac{3}{5-3D})^{\frac{8}{3}-D}(2 \pi)^{D-\frac{5}{3}}(\frac{z}{z_{i}})^{\frac{2}{3}}z^{1-D} [(f_{m})_{w}]^{\frac{8}{3}-D} \ell_{w}^{D-1}\psi_{\epsilon}^{\frac{2}{3}} w_{*}^{2}.
\label{eq 27}\end{equation}

Replacing Eqs. \eqref{eq 26} and \eqref{eq 27} in Eq. \eqref{eq 2}, we obtain the following expression for the Eulerian spectrum of energy normalized by the Eulerian velocity variance (considering $n=0$),

\begin{equation}
E_{w}(0) = \frac{1}{(f_{m})_{w}}\frac{z}{u}.
\label{eq 28}\end{equation}
Replacing $E_{w}(0)$ (given by Eq. 28) and the Eulerian standard deviation of the turbulent wind velocity $\sigma_{w}$, obtained from Eq. \eqref{eq 26}, we obtain the following expression for the vertical asymptotic eddy diffusivity in the convective boundary layer,

\begin{equation}
K(z)=0.2a(\frac{3}{5-3D})^{\frac{3D-11}{3}}(2 \pi)^{\frac{3D-5}{6}}(\frac{z}{z_{i}})^{\frac{11-3D}{6}}z_{i}^{\frac{1-D}{2}} [(f_{m})_{w}]^{\frac{3D-11}{6}-D} \ell_{w}^{\frac{D-1}{2}}\psi_{\epsilon}^{\frac{1}{3}} w_{*}z_{i}.
\label{eq 29} \end{equation}
The constant $a$ is related to the parameter $\beta$ defined in Eq. \eqref{eq 5}. In the literature, we find several values used for this constant \cite{Wandel1962, Hanna1981, Degrazia1998}.   In this work, the value $a=\sqrt{\pi}/4$ will be used \cite{Wandel1962}. \\
In Eq.  \eqref{eq 29}  the following expressions were used for the parameters $(f_{m})_{w}$, $\ell_{w}$ and $\psi_{\epsilon}$ \cite{Degrazia2000}:

\begin{equation}
(f_{m})_{w}= \frac{z}{1.8z_{i}[1-\exp(-4\frac{z}{z_{i}})-0.0003\exp(8\frac{z}{z_{i}})]},
\label{eq 30} \end{equation}

\begin{equation}
\psi_{\epsilon} = [(1-\frac{z}{z_{i}})^{2}(\frac{z}{L})^{-2/3}+0.75]^{3/2},
\label{eq 31} \end{equation}

\begin{equation}
\ell_{w} = 0.25z_{i}(0.01\frac{z_{i}}{L})^{1/2}[1-\exp(-4\frac{z}{z_{i}})-0.0003\exp(8\frac{z}{z_{i}})].
\label{eq 32} \end{equation}

In the next section we will present the fractional dispersion model obtained by Goulart et al. (2020) \cite{Goulart2020}. This model will be used to calculate the contaminant concentration distribution in the planetary boundary layer using the eddy diffusivity given by Eqs. \eqref{eq 11} and \eqref{eq 29}.

\section{The dispersion model}\label{sec3}

An equation for the spatial distribution of the concentration of a non-reactive pollutant in the planetary boundary layer can be obtained by applying of the principle of continuity or conservation of mass, where the flows are represented by the K-theory \cite{Csanady1973, Blackadar1997}. In a Cartesian coordinate system in which the longitudinal direction $x$ coincides with the average wind velocity, the spatial distribution of concentration  $\bar{c}=\bar{c}(x,y,z)$ of a non-reactive substance can be described by the advection-diffusion equation

\begin{equation}\label{eq 33}
 \frac{D \bar{c}}{D t}+\overrightarrow{\nabla}\cdot \overrightarrow{\Pi}_{c}=0,
\end{equation}
where $\frac{D}{Dt}=\frac{\partial}{\partial t}+\vec{u}\cdot \overrightarrow{\nabla}$ is the Lagrangian derivative, $\bar{c}=\bar{c}(x,y,z,t)$ is the average concentration, $\vec{u}$ is the wind velocity, and $\overrightarrow{\Pi}_{c}$ is the concentration flux. A traditional closure for the concentration flux problem given by Eq. \eqref{eq 33} is based on the gradient-transfer approach, which assumes that turbulence causes a net movement of material in the direction of the concentration gradient at a rate that is proportional to the magnitude of the gradient \cite{Pasquill1983}:
\begin{equation}\label{eq 34}
 \overrightarrow{\Pi}_{c}=-K \overrightarrow{\nabla}\bar{c} ,
\end{equation}
where $K$ is the eddy diffusivity.
By replacing Eq. \eqref{eq 34} in Eq. \eqref{eq 33}, considering the steady-state case and choosing a Cartesian coordinate system in which the longitudinal direction $x$ coincides with the average wind velocity, and by neglecting the longitudinal diffusion, we get
\begin{equation}\label{eq 35}
 u\frac{{\partial}\;\overline{c}}{\partial x}-\frac{\partial}{\partial y}(K_{y}\frac{\partial \; \overline{c}}{\partial y})-\frac{\partial}{\partial z} (K_{z}\frac{\partial \; \overline{c}}{\partial z})=0.
\end{equation}
Finally, in order to compare the model with the experimental data found in the literature, the equation for the cross-wind integrated concentration ($\overline{c^{y}}=\overline{c^{y}}(x,z)$) is obtained by integrating Eq. \eqref{eq 35} with respect to $y$ from $-\infty$ to $+\infty$:
\begin{equation}\label{eq 36}
 u\frac{{\partial}\;\overline{c^{y}}}{\partial x}-\frac{\partial}{\partial z}(K_{z}\frac{\partial \; \overline{c^{y}}}{\partial z})=0.
\end{equation}

The Eq. \eqref{eq 36}  represents a classical model used to estimate the concentration distribution of a contaminant in the Earth's atmosphere, obtained by considering closures for concentration flux based on the gradient-transfer approach given by Eq. \eqref{eq 34}.

To consider the pollutant transport as an anomalous diffusion process found in the stable boundary layer, Goulart et al. \cite{Goulart2020} introduced the fractional derivative in the diffusive term  by an appropriate parameterization for the concentration flux and in the advective term, so that Eq. \eqref{eq 36} becomes,
\begin{equation}\label{eq 37}
 u\frac{{\partial^{\alpha}}\;\overline{c^{y}}}{\partial x^{\alpha}}-\frac{\partial}{\partial z}(K_{z}\frac{\partial^{\alpha} \; \overline{c^{y}}}{\partial z^{\alpha}})=0.
\end{equation}

Note that for $\alpha=1$ Eq. \eqref{eq 37} reduces to the classical integer order diffusion equation given by Eq. \eqref{eq 36}.
There are several definitions for the fractional derivative \cite{Metzler2000, Metzler2004, West2014}. In this work, we will use Hausdorf's definition of the fractional derivative \cite{Chen2006},
\begin{equation}\label{eq 38}
\frac{\partial^{\alpha} \overline{c^{y}}} {\partial x^{\alpha}} = \frac{1}{\alpha x^{\alpha-1}}\frac{\partial \overline{c^{y}} }{\partial x}\;\;\;\;\;\text{and}\;\;\;\;\; \frac{\partial^{\alpha} \overline{c^{y}}} {\partial z^{\alpha}} = \frac{1}{\alpha z^{\alpha-1}}\frac{\partial \overline{c^{y}} }{\partial z}
\end{equation}
substituting Eq. \eqref{eq 38} into Eq. \eqref{eq 37}, we get:
\begin{equation}\label{eq 39}
 u \; x^{1-\alpha}\frac{\partial \overline{c^{y}} }{\partial x} = \frac{\partial}{\partial z}\big( K_{z} \; z^{1 - \alpha}\frac{\partial \overline{c^{y}} }{\partial z}\big).
\end{equation}

The fractal dimension D and the alpha value in Eq. \eqref{eq 39} are related by the expression $\alpha=2/(1+D)$ \cite{Aharony1984,Metzler2000,Havlin2002}. In this work, we use the value $D=1.15$. This value is in accordance with the range of values obtained by \cite{Batista2016} for $D$ in the planetary boundary layer,  analyzing the vertical dimension of clouds. As there is no way to deduce an exact value for \textit{D}, we use this value which is found in the literature and refers to turbulence in the atmosphere. The best concentration value (with a small difference between them) was obtained with $D = 1.15$, which is in the range of values found by \cite{Batista2016} for turbulent flow in the planetary boundary layer.

In order for Eq. \eqref{eq 39} to describe a possible real dispersion process in planetary boundary layer, it should be imposed boundary conditions of zero flux on the ground ($z=z_{0}$) and top ($z = h$) and consider that the contaminant is released from an elevated point source with an emission rate of $Q$ at a height of $H_{s}$, i.e.,
\begin{equation}\label{eq 40}
K_{z}\frac{\partial^{\alpha} c^{y} }{\partial z^{\alpha}}=0,\;\;\;\;z=z_{0},\;\;z=h,
\end{equation}
\begin{equation}\label{eq 41}
u c^{y}(0,z)=Q\delta(z-H_{s}),\;\;x=0,
\end{equation}
where $z_{0}$ is the surface roughness length, and $\delta(\cdot)$ is the Dirac delta function.\\
In this work, the Eqs. \eqref{eq 36} and \eqref{eq 39} were numerically solved.

In the next section, the performance of the eddy diffusivity obtained in Eqs. \eqref{eq 11}  and \eqref{eq 29} when used in a fractional Eulerian dispersion model, will be evaluated.

\section{Results and discussion}\label{sec4}

The performance of the \textbf{classical and} fractional derivative models expressed in Eq. \eqref{eq 36} and Eq. \eqref{eq 39}, with the eddy diffusivity given by Eq. \eqref{eq 11} and Eq. \eqref{eq 29} was evaluated against the experimental ground-level concentration using the Prairie Grass and Hanford experiments \cite{Barad1958, Doran1985} for the stable boundary layer and the Copenhagen experiment \cite{Gryning1981} for the convective boun\-dary layer. \\
To compare the simulated and observed concentrations the statistical indices described in Hanna \cite{Hanna1981} are used,
\begin{equation*}
\begin{split}
\text{\textit{NMSE}}\; (\text{normalized mean-square error})&= \frac{\overline{(c_{o}-c_{p})^{2}}}{\overline{c_{o}}\overline{c_{p}}},\\
\text{\textit{Cor}}\; (\text{correlation coefficient})&= \frac{\overline{(c_{o}-\overline{c_{p}})(c_{p}-\overline{c_{p}})}}{\sigma_{o}\sigma_{p}},\\
\text{\textit{FB}}\; (\text{fractional bias})&=\frac{(\overline{c_{o}}-\overline{c_{p}})}{0.5(\overline{c_{o}}+\overline{c_{p}})},\\
\text{\textit{FS}}\; (\text{fractional standard deviations}) &= \frac{\sigma_{o}-\sigma_{p}}{0.5(\sigma_{o}+\sigma_{p})},
\end{split}
\end{equation*}
where $c_{p}$ is the computed concentration, $c_{o}$ is the observed concentration, $\sigma_{p}$ is the computed standard deviation, $\sigma_{o}$ is the observed standard deviation, and the overbar indicates an averaged value. The statistical index \textit{FA}2 represents the fraction of data for $0.5 \leq{c_{p}}/{c_{o}}\leq 2 $. The best results are indicated by values nearest to $0$ in \textit{NMSE}, \textit{FS}, \textit{FB}, and nearest to $1$ in \textit{Cor} and \textit{FA}2. \\
In order to validate the models expressed in Eqs. \eqref{eq 36} and \eqref{eq 39} with the eddy diffusivity given by Eqs. \eqref{eq 11} and \eqref{eq 29}, the experiments of the Prairie Grass and Hanford (stable cases) and Copenhagen (convective case) are described below:
\subsection{Prairie Grass experiment (stable regime)}
The Prairie Grass stable experiment was realized in O'Neill, Nebraska in 1956 \cite{Barad1958}. The pollutant used was $SO_{2}$, emitted from a height of 0.5 m and it was measured by samplers at a height of 1.5 m at five downwind distances (50m, 100m, 200m, 400m and 800m). The Prairie Grass site was flat, with a roughness length of 0.6 cm. The meteorological data for the Prairie Grass stable experiment is shown in Table 1.

\begin{table}[h]
\begin{center}
\caption{Meteorological data for the Prairie Grass experiment (stable case)}\label{tab1}
\begin{tabular}{@{}lcccccc@{}}
\toprule
Exp.   & $h (m)$ &          $u_{*} (ms^{-1})$      &  L(m)   &  $U(ms^{-1} )$                 \\
\midrule
 13     & 23                  & 0.09              & 3.4    & 3.9                        \\
 14     & 12                  & 0.05              & 1.6    & 3.7                        \\
 17     & 131                 & 0.21              & 48     & 3.8                        \\
 18     & 92                  & 0.20              & 25     & 4.0                        \\
 21     & 333                 & 0.38              & 172    & 6.4                        \\
 22     & 400                 & 0.46              & 204    & 7.7                        \\
 23     & 358                 & 0.39              & 193    & 6.5                        \\
 24     & 400                 & 0.38              & 248    & 6.3                        \\
 28     & 81                  & 0.16              & 24     & 3.2                        \\
 29     & 119                 & 0.23              & 36     & 4.3                        \\
 32     & 43                  & 0.13              & 8.3    & 3.6                        \\
 35     & 147                 & 0.24              & 53     & 4.3                        \\
 36     & 36                  & 0.10              & 7.8    & 2.8                        \\
 37     & 216                 & 0.29              & 95     & 5.0                        \\
 38     & 217                 & 0.28              & 99     & 4.8                        \\
 39     & 48                  & 0.14              & 9.8    & 3.6                        \\
 40     & 39                  & 0.11              & 8      & 3.1                        \\
 41     & 117                 & 0.23              & 35     & 4.4                        \\
 42     & 275                 & 0.37              & 120    & 6.3                        \\
 46     & 257                 & 0.34              & 114    & 5.8                        \\
 53     & 54                  & 0.17              & 10     & 4.3                        \\
 54     & 128                 & 0.24              & 40     & 4.5                        \\
 55     & 279                 & 0.37              & 124    & 6.3                        \\
 56     & 194                 & 0.29              & 76     & 5.1                        \\
 58     & 35                  & 0.11              & 6.4    & 3.4                        \\
 59     & 51                  & 0.14              & 11     & 3.4                        \\
 60     & 166                 & 0.28              & 58     & 5.0                        \\
\botrule
\end{tabular}
\end{center}
\end{table}

\newpage

\begin{figure}
\includegraphics[width=1.00\textwidth]{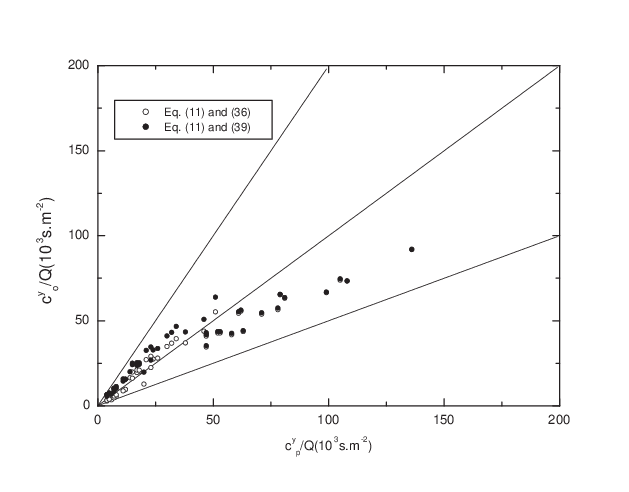}
\centering \caption{Scatter diagram between observed and predicted ($\overline{c^{y}_{o}}$) and predicted ($\overline{c^{y}_{p}}$) for the Prairie Grass experiment } \label{fig1}
\end{figure}

Table 2 shows the results of the statistical indices used to evaluate the model\textsc{\char13}s performance for the stable Prairie Grass experiment. It is possible to see the good performance of the models. Especially in the fractional model (Eq. 39), where the factor of two has a value very close to 1. The results obtained by Kumar and Sharan (2010) \cite{Kumar2010} and Carvalho (2007) \cite{Carvalho2007}, shown in Table 2, were taken directly from their publications. The Lagrangian model (LM) of Carvalho et al. (2007) \cite{Carvalho2007} consist of the solution of the Langevin equation through Picard's Iterative method.

Fig. 1 shows the scatter diagram between the measured and predicted crosswind-integrated concentration using the classical model (Eq. 36) and the fractional model (Eq. 39), using the eddy diffusivity given by Eq. \eqref{eq 11} for the stable Prairie Grass experiment.
\begin{table}[!h]
\centering
\caption{\small Statistical indices of the models given by Eqs. \eqref{eq 36} and \eqref{eq 39} for the stable Prairie Grass experiment.}\label{tab2}
\begin{tabular}{lccccccc}
\hline
Model                                     & \; Cor \;   & \; NMSE \;    & \; FS \;    & \; FB \;      & \; FA2 \; \\
\hline
  Eqs. \eqref{eq 36} and \eqref{eq 11}     & 0.97         & 0.09          & 0.33       & 0.16          & 1.00     \\
 Eqs. \eqref{eq 39} and \eqref{eq 11}      & 0.95         & 0.09          & 0.38       & 0.05          & 1.00 \\
 LM - Carvalho et al. (2007){*} \cite{Carvalho2007}             & 0.81         & 0.39          & -0.18      & -0.057       & 0.95 \\
 EM - Carvalho et al. (2007){**} \cite{Carvalho2007}              & 0.77         & 0.49          & -0.26      & -0.224       & 0.82 \\
 Kumar and Sharan (2010)\cite{Kumar2010}                  & 0.97         & 0.06          & 0.11       & -0.16        & 0.95 \\
  \hline
{*} Lagrangian Model\\
{**} Eulerian Model.
\end{tabular}
\end{table}

\newpage

\subsection{Hanford experiment }
The Hanford experiment was carried out in 1983 in a semi-arid region of south-eastern Washington on generally flat terrain \cite{Doran1985}. Data were obtained from six dual-tracer releases located at 100m, 200m, 800m, 1600m and 3200m from the source during moderately stable to near-neutral conditions. The release height of $SF_{6}$ was 2m and the average release rate was around $0.3gs^{-1}$. The pollutant was collected at a height of 1.5m. The meteorological data observed are given in Table 3.
\begin{table}[h]
\begin{center}
\caption{Meteorological data for the Hanford experiment}\label{tab3}%
\begin{tabular}{@{}lcccccc@{}}
\toprule
Exp.   &  $h (m)$ &           $u_{*} (ms^{-1})$    &  L(m)        &   $U(2 m)(ms^{-1} )$            \\
\midrule
 1      & 325                  & 0.40              & 166          & 3.63                        \\
 2      & 135                  & 0.26              & 44           & 1.42                        \\
 3      & 182                  & 0.27              & 77           & 2.02                        \\
 4      & 104                  & 0.20              & 34           & 1.50                        \\
 5      & 157                  & 0.26              & 59           & 1.41                        \\
 6      & 185                  & 0.30              & 71           & 1.54                        \\
 \botrule
\end{tabular}
\end{center}
\end{table}

\begin{table}[h]
\centering
\caption{\small Statistical indices of the models given by Eqs. \eqref{eq 36} and \eqref{eq 39} for the stable Hanford experiment.}\label{tab4}
\begin{tabular}{lccccccc}
\hline
Model                                     & \; Cor \;   & \; NMSE \;    & \; FS \;    & \; FB \;      & \; FA2 \; \\
\hline
  Eqs. \eqref{eq 36} and \eqref{eq 11}     &    0.90        &   0.14     &    0.23     &    0.08        &    0.97     \\
 Eqs. \eqref{eq 39} and \eqref{eq 11}      &    0.90        &   0.11     &    0.13     &   -0.05        &    0.87 \\
 Kumar and Sharan (2010)\cite{Kumar2010}                  &    0.91        &   0.21     &    0.12     &   -0.14        &    0.73 \\
  \hline
\end{tabular}

\end{table}

\begin{figure}
\includegraphics[width=1.00\textwidth]{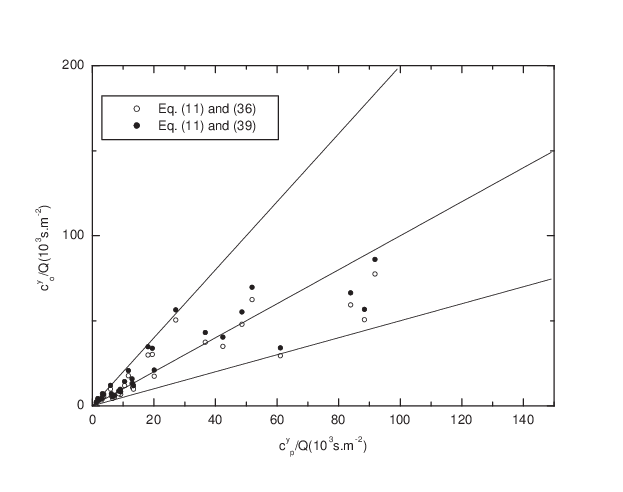}\centering \caption{Scatter diagram between observed and predicted ($\overline{c^{y}_{o}}$) and predicted ($\overline{c^{y}_{p}}$) for the Hanford experiment } \label{fig2}
\end{figure}

Table 4 shows the results of the statistical indices used to evaluate the model's performance for the Hanford experiment. It is possible to see the good performance of the models. In the Kumar and Sharan (2010)  model \cite{Kumar2010}, an eddy difffusivity dependent on the height and distance from the source is used.

Fig. 2 shows the scatter diagram between the measured and predicted crosswind-integrated concentration using the classical model (Eq. 36) and the fractional model (Eq. 39), using the eddy diffusivity given by Eq. \eqref{eq 11} for the Hanford experiment.

{\subsection{Copenhagen experiment }}
 The Copenhagen experiment used tracer $SF_{6}$ data. The tracer was released without buoyancy from a tower at a height of 115m and collected at ground-level position at a maximum of three crosswind arcs of tracer sampling units. The site was mainly residential, with a roughness length of 0.6m. The wind velocity profile used in the air contaminants model has been parameterized following Eq. \eqref{eq 16} where $p=0.09$. The meteorological data observed are given in Table 5.

\begin{table}[h]
\begin{center}
\caption{Meteorological data for the Copenhagen experiment}\label{tab3}%
\begin{tabular}{@{}lccccccc@{}}
\toprule
Exp.   &  $z_{i} (m)$ &           $u_{*} (ms^{-1})$    &  L(m)        &   $U(10 m)(ms^{-1} )$    &  $\sigma_{w} (ms^{-1} )$              \\
\midrule
 1      &   1980                &  0.37             &  46           &  2.1          &    0.83                      \\
 2      &   1920                &  0.74             &  384          &  4.9          &    1.07                     \\
 3      &   1120                &  0.39             &  108          &  2.4          &    0.68                    \\
 4      &   390                 &  0.39             &  173          &  2.5          &    0.47                    \\
 5      &   820                 &  0.46             &  577          &  3.1          &    0.71                     \\
 6      &   1300                &  1.07             &  569          &  7.2          &    1.33                      \\
 7      &   1850                &  0.65             &  136          &  4.1          &    0.87                     \\
 8      &   810                 &  0.70             &  72           &  4.2          &    0.72                     \\
 9      &   2090                &  0.77             &  382          &  5.1          &    0.98                      \\
 \botrule
\end{tabular}
\end{center}
\end{table}

\begin{table}[!h]
\centering
\caption{\small Statistical indices of the models given by Eq. \eqref{eq 36} and Eq. \eqref{eq 39} for the Copenhagen experiment.}\label{tab4}
\begin{tabular}{lccccccc}
\hline
Model                                     & \; Cor \;    & \; NMSE \;     & \; FS \;     & \; FB \;       & \; FA2 \; \\
\hline
  Eqs. \eqref{eq 36} and \eqref{eq 29}     & 0.91         & 0.05           & 0.18         & 0.11           & 1.00 \\
  Eqs. \eqref{eq 39} and \eqref{eq 29}     & 0.90         & 0.04           & 0.14         & 0.04           & 1.00     \\
 Moreira et al. (2005) \cite{Moreira2005}                   & 0.97         & 0.02           & 0.05         & 0.01           & 1.00 \\
 Kumar  et al. (2012)\cite{Kumar2012}     & 0.90         & 0.05           & 0.34         & -0.04          & 0.96 \\
  \hline
\end{tabular}
\end{table}

\begin{figure}
\includegraphics[width=1.00\textwidth]{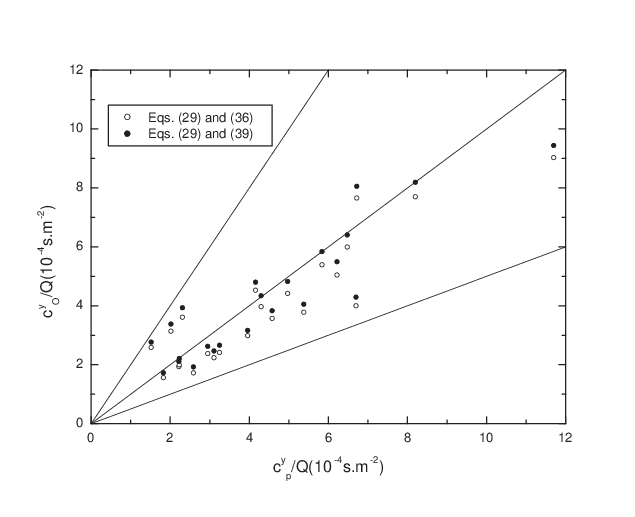}
\centering \caption{Scatter diagram between observed and predicted ($\overline{c^{y}_{o}}$) and predicted ($\overline{c^{y}_{p}}$) for the Copenhagen experiment } \label{fig3}
\end{figure}

Table 5 shows the results of the statistical indices used to evaluate the models performance for the Copenhagen experiment. It is possible to see the good performance of the models.

Fig. 3 shows the scatter diagram between the measured and predicted crosswind-integrated concentration using the classical model (Eq. 36) and the fractional model (Eq. 39), using the eddy diffusivity given by Eq. \eqref{eq 29} for the Copenhagen experiment.
\newpage
\section{Conclusions}\label{sec5}
In this work, we used the $\beta$-model \cite{Frisch1978,Novikov1964} to obtain velocity spectra in the planetary boundary layer in the stable and convective stability regimes. The eddy diffusivity was obtained from the velocity spectra. The Eq. \eqref{eq 11} describes the eddy diffusivity in a stable regime, and Eq. \eqref{eq 29} describes the eddy diffusivity in a convective regime. These eddy diffusivities were used in the classical Eulerian dispersion model, where the derivatives are of integer order (Eq. 36) and in the fractional dispersion model, where the derivatives are of fractional order (Eq. 39) to estimate the distribution of contaminant concentration on the surface of the planetary boundary layer.  The eddy diffusivity proposed in this work considers the anomalous nature of geophysical turbulent flow \cite{Metzler2000, Metzler2004, Goulart2017, Goulart2022}. The results obtained with the fractional (Eq. 39) and classic (Eq. 36) dispersion models using the eddy diffusivity given by Eq. \eqref{eq 11} and Eq. \eqref{eq 29} are satisfactory when compared with the experimental data of the Prairie Grass and Hanford experiments in a stable regime, as indicated in Tables 2 and 4 and Figures 1 and 2 and Copenhagen experiments in a convective regime, as indicated in Table 6 and Figure 3. But a slight superiority in of the performance of the fractional model given by Eq. \eqref{eq 39} is observed in relation to the classic model given by Eq. \eqref{eq 36}. This demonstrates that the use of fractional derivatives, to describe the transport of contaminants in a turbulent flow, where anomalous diffusion occurs, is more adequate than the use of integer derivatives as already observed by Goulart et al. \cite{Goulart2017}. In future work, we intend to use the eddy diffusivity given by Eq. \eqref{eq 11} (stable regime) and Eq. \eqref{eq 29} (convective regime) in more complete fractional dispersion models, which consider the Caputo fractional derivatives and lateral dispersion.

\section*{Acknowledgments}

This work was supported in part by CNPq and CAPES, Brazilian funding agencies.

\section*{Declarations}

\begin{itemize}
\item Funding \\
Not applicable
\item Conflict of interest/Competing interests \\
Not applicable
\item Availability of data and materials \\
The experimental data used in this work were obtained from scientific articles already published and available.
\item Authors' contributions \\
All authors contributed equally to this work
\end{itemize}


\end{document}